\documentclass[prl,twocolumn,aps,showpacs]{revtex4-1}
\newif\ifwithsupp
\withsupptrue   

\usepackage{amsmath}
\usepackage{amssymb}
\usepackage{graphicx}
\usepackage{hyperref}
\usepackage{color}

\def\theEpaps{\ifwithsupp the Appendix\else the supplement\cite{epaps}\fi}

\newcommand{\tr}{\operatorname{tr}}
\def\idty{{\leavevmode\rm 1\mkern -5.4mu I}} 

\def\Rl{{\mathbb R}}

\def\PP{{\mathbb P}}
\def\norm #1{\Vert #1\Vert}

\def\sign{{\mathop{\rm sign}\nolimits}}

\def\braket #1#2{\langle #1 \vert #2\rangle}
\def\ketbra #1#2{\vert #1\rangle \langle #2\vert}
\def\kettbra#1{\ketbra{#1}{#1}}
\def\expect#1{\langle#1\rangle}
\def\tr{\mathop{\rm tr}\nolimits}
\def\abs#1{\vert#1\vert}

\def\Ell{{\mathcal L}}
\def\Cases#1{\left\lbrace\begin{array}{cl}#1\end{array}\right.}

\def\HH{{\mathcal H}}

\def\exit{{\mathcal E}}
\def\dom{{\mathop{\rm dom}\nolimits}\,}
\def\HHh{\widehat\HH}\def\Uh{\widehat U}\def\Hh{\widehat H} 
\def\Th{\widehat T}\def\psih{\widehat\psi}\def\Eh{\widehat E}
\def\Fh{\widehat F}

\def\Pmin{\PP_{\rm min}}

\def\Airy{{\mathop{\rm Ai}\nolimits}}

\begin{document}

\title{Exact Energy-Time Uncertainty Relation for Arrival Time by Absorption}

\author{Jukka Kiukas$^{1}$}
\author{Andreas Ruschhaupt$^{1}$}
\author{Piet O. Schmidt$^2$}
\author{Reinhard F. Werner$^{1}$}
\affiliation{$^{1}$Institut f\"ur Theoretische Physik, Leibniz Universit\"at Hannover, Appelstr. 2, 30167 Hannover, Germany\\
$^{2}$ QUEST Institute for Experimental Quantum Metrology,
Physikalisch-Technische Bundesanstalt and Leibniz University Hannover,
Bundesallee 100, 38116 Braunschweig}

\begin{abstract}We prove an uncertainty relation for energy and arrival time, where the arrival of a particle at a detector is modeled by an absorbing term added to the Hamiltonian. In this well-known scheme the probability for the particle's arrival at the counter is identified with the loss of normalization for an initial wave packet. Under the sole assumption that the absorbing term vanishes on the initial wave function, we show that $\Delta T\Delta E\geq \sqrt p\hbar/2$ and $\expect T\Delta E\geq 1.37\sqrt p\hbar$, where $\expect T$ denotes the mean arrival time, and $p$ is the probability for the particle to be eventually absorbed. Nearly minimal uncertainty can be achieved in a two-level system, and we propose a trapped ion  experiment to realize this situation.
\end{abstract}


\maketitle
\ifwithsupp\else\nocite{epaps}\fi 

\section{Introduction}
From Heisenberg's seminal 1927 paper \cite{Hei}, uncertainty relations have been recognized as a fundamental feature of quantum mechanics. Heisenberg gives a semi-classical heuristic discussion and his ``uncertainties'' are conceptually different in different parts of his paper. Accordingly, they are not given a precise quantitative  meaning in the mathematical language of quantum theory. On the other hand, modern textbooks all agree on ``the''
uncertainty relation, namely a version stated and proved by Kennard \cite{Ken} in the same year Heisenberg's paper appeared. Kennard achieved an important clarification both conceptually, by defining the uncertainties as the standard deviation of operationally well-defined probability distributions, and also quantitatively, so it becomes possible to say that some experiment realizes an uncertainty product within 3\% of the absolute minimum. This clarification was so successful that other aspects of Heisenberg's paper, like his discussion of the precision of a position measurement by microscope versus the momentum disturbance by the measurement, fell into disrepute. Nevertheless, these ideas are not only heuristically meaningful, but can be made operationally precise and proved as theorems in the quantum formalism \cite{Wer04,BuschReviewQP}. Another group of sharp results looks at the position and momentum distributions like Kennard, but defines the ``spread'' in a different way. Particularly interesting is a non-parametric version in terms of entropies \cite{MaassenUff,WehnerSurvey,OppenheimScienceReport}, which has applications in cryptography \cite{UncertCrypto}. Since many experiments are nowadays approaching quantum limits, it is perhaps becoming more important to think of uncertainty relations not as a single fact, but as a circle of ideas in which traditional heuristic interpretations coexist with an increasing family of exact results.

The ambiguities in the literature multiply when we go from position-momentum uncertainty to energy-time uncertainty. In most cases, energy-time uncertainty relations are invoked in a handwaving fashion only. There are very few
conceptually clear and quantitatively meaningful formulations (see \cite{BuschReview} for a review).  For one kind of arrival time observables, the so-called covariant observables \cite{Kijowski,HolevoUncert,Wer86}, a Kennard-like uncertainty relation is well known: Then $\Delta E$ is the standard deviation of the energy observable and $\Delta T$ the standard deviation of arrival time. However, this notion of arrival times is rather inflexible and makes no sense for finite dimensional systems. These drawbacks can be cured in a more realistic theory of arrival times \cite{All69,Wer87}, in which the counter is modeled by an absorbing term in the Hamiltonian.

The purpose of this paper is to provide a sharp quantitative formulation of uncertainty for energy and absorptive arrival time.
The possibility of such an uncertainty relation is somewhat surprising (and totally missed in the literature, including \cite{Wer87}), since the dynamics including the counters, which is used to define $\Delta T$, is not generated by the Hamiltonian used to define $\Delta E$, so there would appear to be no universal trade-off inequality.
Nevertheless, we will prove such a relation under only a mild and natural assumption on the initial state.
An immediate advantage of the absorptive arrival time approach is that it also applies to certain finite dimensional models used in quantum optics. Thus, we propose a concrete minimal uncertainty experiment with trapped ions.

Our paper is organized as follows. After a brief introduction to absorptive arrival times, we will state the
relation and the conditions for near-equality. The main ideas of the proof are
sketched in the following section, with a full proof given in
\theEpaps. We then show that (very nearly) minimum uncertainty can be realized in a standard quantum optical setting.

\section{Arrival Times}
Consider a quantum system with Hilbert space $\HH$ and Hamiltonian $H$. Starting from some initial state $\psi\in\HH$, we would like to determine the probability
distribution of arrival times at some counter. There are different approaches to this problem, varying in the degree of detail with which the counter is described.

The coarsest, and simplest, description focuses just on transformation behavior:
starting from the time-evolved state $\psi_t=\exp(-iHt/\hbar)\psi$ we should get the same arrival time distribution, but shifted by $t$.
{\em Covariant arrival observables} with this property have been studied extensively \cite{Kijowski,MugainBook,Wer86},
and satisfy a general energy-time uncertainty relation \cite{HolevoUncert}.
However, transformation behavior alone is not sufficient to single out a convincing model for a given experimental counter array.
Moreover, this approach requires the Hamiltonian to have a purely continuous spectrum and is hence limited to infinite dimensional Hilbert spaces.

At the other extreme we can make a detailed {\em detector model}, e.g., by interaction with an ionizable atom \cite{RuschhauptinBook}. The drawback of this scheme is that, although undoubtedly physically correct, the interacting system is very hard to treat, and a disproportionate amount of the analysis of a given experiment would go just into the detection process.

The {\em absorptive arrival times} approach adopted in this paper is a good compromise between these extremes: It was first worked out in detail in \cite{All69} and describes the detector by a non-hermitian term $-iD$ added to the Hamiltonian $H$, which is thus replaced by $K=H-iD$. Thereby the unitary time evolution operator $U_t=\exp(-iHt/\hbar)$ is modified to a semigroup of contractions, i.e., operators $B_t=\exp(-iKt/\hbar)$ ($t\geq0$) such that $\norm{B_t}\leq1$.  We now interpret the loss of normalization, i.e., $1-\norm{B_t\psi}^2$, of a quantum state $\psi$ as the probability that the particle did arrive before time $t$. More generally, for any time interval $[t,s]$ with $0\leq t\leq s$ the probability of arrival in that interval is given by the expectation of the operator
\begin{equation}\label{Ftt}
    F([t,s])= B_t^*B_t-B_s^*B_s.
\end{equation}
Thus, given a quantum state $\psi$, we get a probability density $\PP(t)$ on the positive time axis by
\begin{eqnarray}\label{time-distr}
     \PP(t)&=&-\frac1p\ \frac d{dt} \braket{B_t\psi}{B_t\psi}, \\
                  p&=& 1-\lim_{t\to\infty}\norm{B_t\psi}^2=\braket\psi{(\idty- R)\psi}.
\end{eqnarray}
Here the limit $R=\lim_{t\to\infty}B_t^*B_t$ exists because it is over a decreasing family of positive operators, and $p$ is the total absorbtion probability.
For real detectors not every absorption actually leads to a detected click. Suppose the probability for this process is $q$. Then we would observe a click in $[t,t+dt]$ with probability $qp \PP(t)dt$. Hence $\PP(t)$ can be determined from the experimental data by normalizing the observed click distribution, independently of $q$ or $p$. The important distinction between these two probabilities is that the process leading from absorption to detection is independent of the particle dynamics and introduces no change in the further quantum evolution of the particle. Therefore, $q$ cannot enter in the uncertainty relation, but as we will see, $p$ does.

We denote by $\expect T$ and $\expect{T^2}$ the first and second moment
of the probability distribution $\PP(t)$ after \eqref{time-distr}, and set
$(\Delta T)^2={\expect{T^2}-\expect{T}^2}$.
Defining $(\Delta E)^2=\braket{\psi}{H^2\psi}-{\braket{\psi}{H \psi}}^2$,
we can hence look for a universal lower bound on the product $\Delta T\cdot\Delta E$.

Without further conditions such a lower bound cannot hold. Indeed, $\Delta T$
can be computed knowing $K$ and $\psi$, whereas $\Delta E$ depends on $H$ and
$\psi$. For example, if we now set $K=H-i\alpha\idty$, we get $p=1$ and $\PP(t)=2\alpha
e^{-2\alpha t}$ independently of $\psi$. Clearly, this cannot imply any
constraint on the energy distribution. Similarly, if the initial state is a
joint eigenvector of $H$ and $K$, then $\Delta E=0$ and $\Delta T$ has a
finite value belonging to an exponential distribution.

\section{Uncertainty Relation}

From the last paragraph it seems that it makes no sense to look for a general time-energy uncertainty relation in this setting.
However, as we will now show, a simple and physically natural condition suffices to derive one. Loosely speaking,
the condition is that the initial wave function has no overlap with the detector. More
formally, if $D$ describes the detector as explained above, we want $D\psi=0$,
or $H\psi=K\psi$. Since these operators are usually unbounded, we also have to
specify the domains. Writing $\dom X$ for the domain of the operator $X$, we
require that
\begin{equation}\label{hh0}
    \psi\in\dom K^2\cap\dom H
     \quad\mbox{and}\ H\psi=K\psi\Bigr..
\end{equation}
The main result of our paper is that under this condition
\begin{equation}\label{ETU}
    \Delta T\cdot\Delta E > \frac\hbar2\ \sqrt p\quad.
\end{equation}
The dependence on $p$ expresses the fact that for a small detection operator $D$ only a few particles are ever detected ($p\approx0$), so observing arrival times cannot imply a strong constraint on $\Delta E$. The power $\sqrt p$ is explained in the sketch of proof below.

The arrival time distribution $\PP(t)$ is always supported by the positive time axis $\Rl_+$. Therefore, the mean arrival time $\expect T$ is always positive, and can take the place of $\Delta T$ in the uncertainty relation. Often $\expect T$ is of more immediate relevance than $\Delta T$. Consider, for example, the decay of a metastable state. The initial wave function is trapped inside a potential barrier, through which it will eventually tunnel. A detector is placed at a distance from the potential (so our initial condition $K\psi=H\psi$ is satisfied). Then $\expect T$ is directly an expression of the lifetime of the metastable state. Under the same conditions as for \eqref{ETU} we prove that
\begin{equation}\label{EeTU}
    \expect T\cdot\Delta E \geq C\,\hbar\ \sqrt p\ ,
\end{equation}
where
  $C=2(-Z_1/3)^{(3/2)}\approx1.376
  $
is a numerical constant involving the first negative zero $Z_1$ of the Airy function.

Cases of (nearly) minimal uncertainty are in both cases connected to specific probability distributions $\Pmin(t)$. Let $(1+\varepsilon)$ denote the ratio of the left hand side to the right hand side in \eqref{ETU} or \eqref{EeTU}.  Then, for a suitable scaling factor $\lambda$ and shift $\tau$,
\begin{equation}\label{nearmin}
    \int\!\!dt \Bigl|{\PP(t)-\lambda\Pmin(\lambda(t-\tau))}\Bigr|
        \leq \gamma\sqrt\varepsilon
\end{equation}
For \eqref{ETU}, $\Pmin(t)$ is Gaussian, and $\gamma=\sqrt2$, as for the standard position-momentum uncertainty relation. For \eqref{EeTU}, $\Pmin(t)$ is the square of the Airy function, and $\gamma=1.888$. We remark that no such conclusion can be drawn for the energy distribution. In fact, although \eqref{ETU} is nearly sharp for the two-level system discussed below, the energy distribution is concentrated on two points, and is hence very different from a Gaussian. Another consequence is that equality (i.e., $\varepsilon=0$) is impossible in \eqref{ETU}, because every Gaussian has a tail extending to the negative half axis. Equality is possible for \eqref{EeTU}, however.

\section{Sketch of proof}
The basic idea of the proof is a so-called dilation construction, by which the
system is transformed in such a way that energy and time become conjugate
self-adjoint operators, and theorems about position-momentum pairs can be
used. The full argument allowing also unbounded $H$ and $K$ is given in
\theEpaps. Here we give a rough version in theoretical physics (rather than
mathematical) style. For simplicity we focus on the $\Delta E\,\Delta T$
relation \eqref{ETU}, and assume that $K=H-iD$ for a bounded positive operator
$D$. For the dependence on the total absorption probability $p$ we only
give an argument showing why the square root is the expected power. For the
main part of the proof we then just assume that $p=1$ for all initial
states. This is true, e.g., in typical finite dimensional systems.

To understand the $\sqrt p$ factor in \eqref{ETU} and \eqref{EeTU}, consider
some $\HH,H,K,\psi$ as above, with full absorption $p=1$.
Compare it with the following modification,
which just adds a part of the Hilbert space that is ``not seen'' by the
detector. Formally, the Hilbert space $\HH$ is enlarged by an additional orthogonal
summand, on which $K$ vanishes, and which contains an eigenvector $\phi_0$ of
$H$ with $\braket{\phi_0}{H\phi_0}=\braket\psi{H\psi}=E$. In this system we
consider the initial vector $\psi'=\sqrt p'\psi+\sqrt{1-p'}\phi_0$, where $0\leq
p'\leq1$. This $p'$ is precisely the absorption parameter for the extended system.
The time distribution does not change at all since we defined it as normalized by $p'$.
The energy expectation also
does not change, but $\Delta_{\psi'}
E^2=p'\Delta_\psi E^2+(1-p')(\Delta_{\phi_0}E)^2=p'\Delta_\psi E^2$.
Hence \eqref{ETU} and \eqref{EeTU}, which hold with $p=1$ for $\psi$,
are equivalent to their versions for $\psi'$ with the factor  $\sqrt{p'}$.
More complex ways in which the particle might
not be seen by the detector are covered by the full proof. From now on we just
assume $p=1$.

We will associate with any wave function $\psi\in\HH$ another wave function $\psih$, which is a function of time, so that $\abs{\psih(t)}^2$ is the arrival probability density. In other words, $\psih$ is a wave function in a {\em time representation}. For each $t$, $\psih(t)$ lies in the original Hilbert space $\HH$. We define $\psih=J\psi$ with the linear operator
\begin{equation}\label{psih0}
    (J\psi)(t)=\Cases{ \sqrt{2/\hbar}\,D^{1/2}\,B_t\psi&\mbox{if\ }t\geq0\\0&\mbox{otherwise}},
\end{equation}
where the square root of the positive operator $D$ is taken in the operator sense. Then, for $t\geq0$,
\begin{eqnarray}\label{psiht2}
    \norm{(J\psi)(t)}^2&=&\frac2{\hbar} \braket{B_t\psi}{DB_t\psi} \nonumber\\
                     &=&- \frac d{dt}\braket{B_t\psi}{B_t\psi} = \PP(t)\ .
\end{eqnarray}
Here we used the assumption that $p=1$ for all states, so the normalizing factor $p$ in \eqref{time-distr} can be omitted.
Equivalently, $J^*J=\idty$. We denote by $\Th$ the self-adjoint operator of multiplication by $t$. Then the required moments of $\PP(t)$
are $\expect T=\braket{\psih}{\Th \psih}$ and $\expect{T^2}=\braket{\psih}{\Th^2 \psih}$.

The translation in the time representation is generated by the operator
$\Hh=i\hbar\,d/(dt)$. By differentiating \eqref{psih0} and using the assumption $H\psi=K\psi$, we find, for $t\geq0$,
\begin{eqnarray}\label{Hpsih0}
    (\Hh J\psi)(t)&=&i\hbar\,\sqrt{2/\hbar}\,D^{1/2}\,B_t(-iK/\hbar)\psi \nonumber\\
                  &=&(JK\psi)(t)=(JH\psi)(t).
\end{eqnarray}
This obviously also holds for $t<0$. At $t=0$ the definition \eqref{psih0} would appear to allow a jump discontinuity, leading to a $\delta(t)$ contribution in
\eqref{Hpsih0}. But since $D\psi=0$ we also have $(J\psi)(0)=0$, and hence there is no jump. We can thus take \eqref{Hpsih0} as an equation for functions, i.e., $\Hh\psih=\Hh J\psi=JH\psi$. Therefore, using $J^*J=\idty$, the variances $\Delta E^2=\norm{H\psi}^2-\braket\psi{H\psi}^2$ and the corresponding $\Delta \widehat E^2=\norm{\Hh\psih}^2-\braket\psih{\Hh\psih}^2$  are the same. The time-energy uncertainty relation \eqref{ETU} thus follows from the standard position-momentum one, applied to ``position'' $\Th$ and ``momentum'' $\Hh$.

It is well-known that the uncertainty relation can be derived from the operator inequality $\Hh^2+\Th^2\geq\idty$ and a dimensional scaling relation. A similar argument applied to wave functions only on the positive time axis, using the inequality $\Hh^2+\Th\geq\lambda\idty$ gives \eqref{EeTU}. $\lambda$ in this equation
is determined from the ground state problem of a particle in a linear potential in front of a wall, which is solved in terms of the Airy function. Using the gap to the first excited state in this problem (and the oscillator Hamiltonian) gives the estimates for near-equality \eqref{nearmin}.

\section{Emission from a two-level system}
Let us consider the simplest possible system to which the relations
(\ref{ETU}) and (\ref{EeTU}) apply. Unlike the covariant observable approach, in which the spectrum of $H$ has to be continuous, finite dimensional systems are included. We consider a two-level system with
\begin{equation}\label{HK2}
   H=\frac{\hbar}{2}\begin{pmatrix}0&\Omega\\\Omega&0\end{pmatrix}\quad,\quad
   D=\frac{\hbar}{2}\begin{pmatrix}0&0\\0&\gamma\end{pmatrix}
\end{equation}
where $\Omega,\gamma>0$ are parameters. The relevant quantity is $\gamma/\Omega$, since we could make $\Omega/2=\hbar=1$ by a choice of units.
As the initial state we take $\psi=(1,0)^T$, so our basic assumption
$H\psi=(0,\hbar\Omega/2)^T=(H-i D)\psi=K\psi$ is satisfied. Obviously,
$\expect H=0$ and $\expect{H^2}=\frac{\hbar^2}{4}\Omega^2$, so $\Delta
E=\frac{\hbar}{2}\Omega$.

\begin{figure}[t]
  \includegraphics[width=0.8\linewidth]{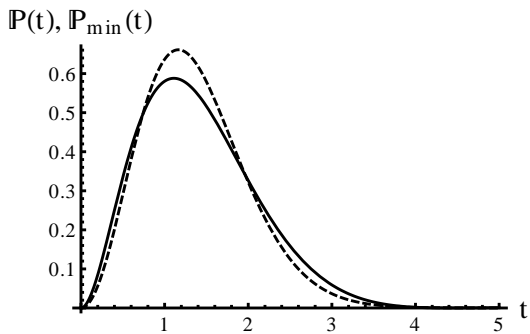}
  \caption{Thick curve: Arrival time probability density $\PP(t)$ after \eqref{time-distr}, for $\Omega/2=\hbar=1$ and $\gamma=2\sqrt2$.
     Dashed: probability distribution $\Pmin (t)$.  }
  \label{Fig:22bump}
\end{figure}

One can explicitly exponentiate $B_t=\exp(-i K t/\hbar)$, and hence compute the probability density $\PP(t)$.
The result is shown in Fig.~\ref{Fig:22bump}.
The moments are also readily calculated.
Both $\expect T$ and $\Delta T$ attain their minimum when
$\gamma=\sqrt2 \Omega$. Then the uncertainty inequalities are satisfied  as
$1/{\sqrt2}\approx 0.707>0.500$ for \eqref{ETU},  and
$\sqrt2\approx 1.414>1.376$ for \eqref{EeTU},
so $\expect T \Delta E$ reaches the minimum to within 3\%! This implies by \eqref{nearmin} that the arrival time distribution $\PP(t)$ must be close to $\Pmin(t)$. The comparison is shown in  Fig.~\ref{Fig:22bump}.

\section{Experimental implementation}
\begin{figure}[ht]
  \includegraphics[width=0.8\linewidth]{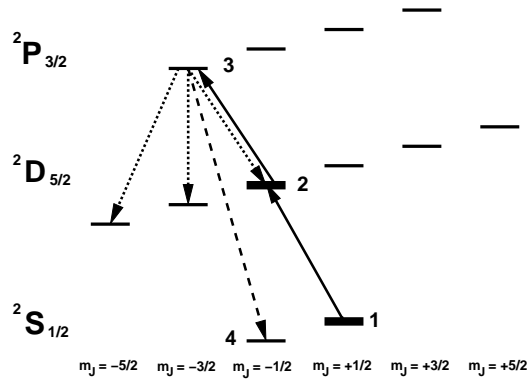}
  \caption{Level scheme of $^{40}$Ca$^+$, the relevant levels 1,2 for a near-minimum
    uncertainty experiment are emphasized, the transitions with their branching ratios $\beta$ are:
    $\lambda_{1\to 2} = 729.147$ nm, $\Gamma_{2\to 1}=2\pi \times 0.0544$~Hz, $\beta_{2\to 1}=1$;
    $\lambda_{2\to 3} = 854.209$ nm, $\Gamma_{3\to 2}=2\pi \times 105$~kHz, $\beta_{3\to 2}=6.8\%$;
    $\lambda_{4\to 3} = 393.366$ nm, $\Gamma_{3\to 4}=2\pi \times 21.2$~MHz, $\beta_{3\to 4}=92\%$.}
  \label{Fig:Ca}
\end{figure}

Instances of the uncertainty inequality as such are not an interesting
experimental target, since it is impossible {\it not} to implement it. On the
other hand, instances with near minimal uncertainty can be a challenge.
The two-level scheme approximations described above naturally arises in quantum optics
when looking at an atom interacting with lasers.
A feasible level scheme is the one of a single $^{40}$Ca$^+$ atom (similarly level schemes exist in other atoms), trapped in a Paul trap, which is shown in
Fig.~\ref{Fig:Ca}. The transitions $1\to 2$ and $2\to 3$ are driven by
on-resonance lasers resulting in Rabi-frequencies $\Omega_{12}$
resp. $\Omega_{23}$. The narrow quadrupole transition $1\to 2$ can be frequency resolved by a narrow-linewidth laser, whereas the $2\to 3$ transition is selected by choosing $\sigma$-polarized light. If $|\Omega_{23}| \ll \Gamma_{3\to4}$, we will
get approximately an efficient decay rate
$\gamma = |\Omega_{23}|^2/\Gamma_{3\to 4}$ for level 2.
If we consider only levels $1$ and $2$, the corresponding effective Hamiltonian,
within the rotating wave approximation and in the interaction picture to get
rid of any time dependence, will be $K=H-iD$ as in \eqref{HK2}, with this $\gamma$ and $\Omega=\Omega_{12}$.
The initial level $1$ can be efficiently prepared via optical pumping. The arrival time in state $2$ is then measured
by measuring the time of the first spontaneously emitted photon on the $3\to 4$ transition.
In the proposed scheme, the total absorbtion probability $p$ is smaller than $1$ due
to spontaneous scattering events from state 3 back to state 2. However, these
events are suppressed by more than a factor of 200 due to favorable branching
ratios and small Clebsch-Gordan coefficients on the $2\to 3$ transition. All
other spontaneous decay channels result merely to a reduced detection
efficiency and do not affect the uncertainty relation. A typical Rabi
frequency of $\Omega_{12}\sim 2\pi\times 100$~kHz would require an easily
achievable $\Omega_{23}\sim2\pi\times 1.73$~MHz to reach approximately the
minimum of the product $\expect T \Delta E$, see Fig. \ref{Fig:22bump}.
The required time resolution for the photon detector in this scenario is on
the order of a few ns, well within reach of current technology. Alternative
implementations include vacuum-stimulated Raman transitions in an atom
strongly coupled to a leaky cavity \cite{CavityExperiment}.

\section{Outlook}
The approach given here can be applied to other situations involving time in quantum mechanics, such as tunneling and decay scenarios. A standard detector model for such applications is the pointlike counter, formally given by $D=\lambda\delta(x)$. In the finite dimensional ($n>2$) setting it remains an interesting problem to characterize the arrival time densities $\PP(t)$, which can be engineered by a suitable choice of laser couplings and initial states.

\vskip12pt

\section{Acknowledgements}
We acknowledge funding from the BMBF (Ephquam project), the EU (projects
CORNER and COQUIT), and the Emil Aaltonen Foundation. P.S.
acknowledges support from the Cluster of Excellence QUEST,
Hannover, and the Physikalisch-Technische Bundesanstalt, Braunschweig.

\nocite{Muga1,Muga2}

\ifwithsupp\section*{Appendix: Proof}\else\section*{Proof}\fi\label{sect:proof}
The formal proof of the statements in the paper follows the sketch given there, with the following main differences:
\begin{itemize}
\item
The operators $K$ and $H$ are allowed to be unbounded, and domain questions are treated explicitly. This covers cases, where
$K-H$ is not meaningful as an operator (e.g., a $\delta$-function potential). Therefore, the difference $D=i(K-H)$ appears nowhere, and the dilation operator $J$ has to be defined differently from \eqref{psih0}.
\item There is no assumption of total absorption, i.e., $p<1$ is allowed.
\item The differentiation \eqref{Hpsih0} is only carried out under suitable scalar products, avoiding the discussion distributional derivatives.
\item We set $\hbar=1$ for notational convenience.
\end{itemize}

\subsection{The Dilation}
We begin by recapitulating the dilation construction from \cite{Wer87}. Since $B_t$ is a contraction semigroup, $-d/dt\,\braket{B_t\psi}{B_t\phi}$ is a positive semidefinite sesquilinear form on $\dom K$. We denote its completion by $\exit$, which we call the exit space, and by $j:\dom K\to\exit$ the embedding map, so that
for $\phi,\psi\in\dom K$:
\begin{eqnarray}\label{jj}
    \braket{j\psi}{j\phi}_\exit&=&\left.-\frac d{dt}\braket{B_t\psi}{B_t\phi}\right\vert_{t=0^+}
                 \nonumber\\
                 &=&i\left(\braket\psi{K\phi}-\braket{K\psi}\phi\right)\ .
\end{eqnarray}
The basic dilation operator $J$ maps into the space
\begin{equation}\label{HHh}
    \HHh=\Ell^2(\Rl,dt;\exit),
\end{equation}
which is the space of ``$\exit$-valued wave functions'' $t\mapsto\Psi(t)\in\exit$ such that $\norm\Psi^2=\int_{\infty}^\infty \!\!dt\,\norm{\Psi(t)}_\exit^2<\infty$. For $\psi\in\dom K$, we define
\begin{equation}\label{JJ}
    \bigl(J\psi\bigr)(t)=\Cases{j(B_t\psi)&\mbox{if\ }t\geq0\\ 0&\mbox{if\ }t<0}.
\end{equation}
Then
\begin{eqnarray}\label{jjnorm}
    \norm{J\psi}^2&=&\int_0^\infty\!\!dt\ \braket{jB_t\psi}{jB_t\psi}
                   \nonumber\\
                &=&\int_0^\infty\!\!dt\ \left(-\frac d{dt}\braket{B_t\psi}{B_t\psi}\right)
                   \nonumber\\
                &=&\Bigl.-\norm{B_t\psi}^2\Bigr\vert_{t=0}^{\infty}
                   \nonumber\\
                &=&\braket\psi{(\idty-R)\psi}, \\
    \mbox{where}\quad
             R&=&\lim_{t\to\infty}B_t^*B_t\ .
\end{eqnarray}
Hence $\norm{J\psi}\leq\norm\psi$, and $J$ extends by continuity to a unique operator $J:\HH\to\Ell^2(\Rl,dt;\exit)$, which we will denote by the same letter.

\subsection{Time Distributions}
Now in the dilation space $\HHh$ we have a time operator $\Th$, which acts by multiplication with $t$. Its spectral projections $\Fh([t,s])$ are a dilation of the arrival time observable $F$. That is, in analogy with \eqref{jjnorm} we get, for $0<t<s$ and $\phi,\psi\in\dom K$:
\begin{eqnarray}
  \braket{J\phi}{\Fh([t,s])J\psi}
                &=&\int_t^s\!\!d\tau\ \braket{jB_\tau\phi}{jB_\tau\psi}
                   \nonumber\\
                &=&-\Bigl.\braket{B_\tau\phi}{B_\tau\psi}\Bigr\vert_{\tau=t}^s
                   \nonumber\\
                &=&\braket\phi{F([t,s])\psi}\ .
\end{eqnarray}
Since the operators involved are bounded and $\dom K$ is dense, this is summarized in the operator equation \begin{equation}\label{Fdilate}
    F([t,s])=J^*\Fh([t,s])J.
\end{equation}
In particular, the probability density $\PP$ after \eqref{time-distr} is equal to $\PP(t)=\norm{(J\psi)(t)}^2/p$, i.e., the probability density associated to the time operator $\Th$ by the vector
\begin{equation}\label{rhoh}
    \psih=\frac1{\sqrt p}\, J\psi.
\end{equation}
This vector is normalized, because $J^*J=\idty-R$, so $p=\braket{\psi}{J^*J\psi}$.

\subsection{Energy Distributions}
The dilated energy operator $\Hh$ is defined as the generator of the translations $\Uh_t=\exp(-i\Hh t)$, where
\begin{equation}\label{Uh}
    \bigl(\Uh_\tau\Psi\bigr)(t)=\Psi(t+\tau)
\end{equation}
for $t,\tau\in\Rl$. Obviously, for $t,\tau\geq0$ and $\psi\in\dom K$, we have $(\Uh_\tau J\psi)(t)=
j(B_{t+\tau}\psi)=(JB_\tau\psi)(t)$, and hence $J^*\Uh_tJ=J^*JB_t$. By taking adjoints we also get an expression for $t<0$. To summarize,
\begin{equation}\label{JUhJ}
  J^*\Uh_tJ=\Cases{(\idty-R)B_t&\mbox{if\ } t\geq0\\
                  B_{-t}^*(\idty-R)&\mbox{if\ } t\leq0}.
\end{equation}
Again, these expressions are directly verified on $\dom K$ and extended by continuity to all of $\HH$.
This shows that in the special case $R=0$, $\Uh$ is a dilation of the semigroup $B_t$ in the sense of Sz.-Nagy \cite{NagyFoias,Davies}. Since $B_t^*RB_t=R$, we can commute generators and $R$ in the sense that
\begin{eqnarray}
   R\,\dom K&\subset&\dom K^*\ , \nonumber\\
   RK&=&K^*R\ .
\end{eqnarray}
The $\Hh$-distribution with respect to $\psih$ from \eqref{rhoh} is the measure $\braket\psih{\Eh(d\omega)\psih}$, where $\Eh$ is the spectral measure of $\Hh$. The characteristic function (Fourier transform) of this distribution is
\begin{eqnarray}\label{cfunc}
    C(t)&=&\int_{-\infty}^\infty\!\!e^{-i\omega t}\braket\psih{\Eh(d\omega)\psih}=\braket\psih{\Uh_t\psih}
            \nonumber\\
        &=& \frac1p\ \braket\psi{J^*\Uh_tJ\psi},
\end{eqnarray}
which can be evaluated further using \eqref{JUhJ}. The moments of the $\Hh$-distribution are obtained by differentiating $C$ at zero. In general, of course, the piecewise defined function \eqref{JUhJ} is not differentiable at zero. However, with our assumption on $\psi$, we can establish that the right and left derivatives coincide.

The conditions  $\psi\in\dom K\cap\dom H$ and $K\psi=H\psi$ imply that $\norm{j(\psi)}^2=i(\braket\psi{H\psi}-\braket{H\psi}\psi)=0$, so $j\psi=0$ for such $\psi$. Moreover, for all $\phi\in\dom K$,
\begin{equation}
    \braket\psi{K\phi}-\braket{K\psi}\phi=-i\braket{j\psi}{j\phi}=0,\nonumber
\end{equation}
whence $K^*\psi=K\psi=H\psi$. Now the first derivative of the characteristic function becomes
\begin{eqnarray}\label{firstorderpsi}
   ip\, C'(0)&=&\Cases{\braket\psi{(\idty-R)K\psi}&\mbox{for\ } t=0^+\\
                                    \braket{\psi}{K^*(\idty-R)\psi}&\mbox{for\ } t=0^-}
\\\nonumber
        &=&\braket\psi{(\idty-R)H\psi}=\braket{H\psi}{(\idty-R)\psi}.
\end{eqnarray}
Similarly, for the second order, and with the additional property $\psi\in\dom K^2$ we get
\begin{eqnarray}\label{secondorderpsi}
   -p\, C''(0)&=&\Cases{\braket\psi{(\idty-R)K^2\psi}&\mbox{for\ } t=0^+\\
                                    \braket{\psi}{(K^*)^2(\idty-R)\psi}&\mbox{for\ } t=0^-}
        \nonumber\\
        &=&\braket{H\psi}{(\idty-R)H\psi}.
\end{eqnarray}
Hence, with $\lambda=\braket{\psi}{H\psi}$ we get the variance
\begin{eqnarray}\label{variancepsih}
  (\Delta E)^2&=& \braket{(H-\lambda\idty)\psi}{(H-\lambda\idty)\psi} \nonumber\\
    &\geq&\braket{(H-\lambda\idty)\psi}{(\idty-R)(H-\lambda\idty)\psi} \nonumber\\
    &=& p (-C''(0)-2i\lambda C'(0)+\lambda^2C(0))\nonumber\\
    &=& p \int(\omega-\lambda)^2\tr \braket\psih{\Eh(d\omega)\psih} \nonumber\\
    &\geq& p \min_\lambda\int(\omega-\lambda)^2\tr \braket\psih{\Eh(d\omega)\psih} \nonumber\\
    &=& p (\Delta\Hh)^2 \nonumber.
\end{eqnarray}
To summarize:
\begin{equation}\label{DelEHh}
   \Delta E \geq\sqrt p\ \Delta\Hh.
\end{equation}
In particular, $\Delta E\Delta T\geq\sqrt p\,\Delta\Hh\Delta\Th\geq\sqrt p\,/2$, which proves \eqref{ETU}.
However, since we also want \eqref{EeTU} and the error estimate \eqref{nearmin}, we need a more detailed analysis.

\subsection{Conversion to ground state problems}
The inequalities \eqref{ETU} and \eqref{EeTU} are for a product of moments. It is advantageous to turn this into inequalities for the expectations of a {\em sum} of operators. For this we use a standard trick, which basically amounts to a dimensional analysis, and hence assures that we estimate quantities of the physical dimension time$\times$energy. If we change the unit of time, we transform $\Th\mapsto\eta^{-1}\Th$, and $\Hh\mapsto\eta\Hh$. This change is also implemented by a unitary operator in $\HHh$. Similarly, we can shift $\Hh$ and $\Th$ by multiples of the identity with unitary operators. This is the idea behind choosing the operators
\begin{eqnarray}\label{AoperatorsPaps}
   X&=&\eta^2(\Hh-\varepsilon)^2+\eta^{-2} (\Th-\tau)^2, \\
   Y&=&\eta^2(\Hh-\varepsilon)^2+\eta^{-1} \Th,
   \label{AoperatorsPops}
\end{eqnarray}
where $Y$ is considered as a quadratic form on the subspace of functions vanishing for $t<0$. These are just standard Schr\"odinger operators with $\Hh$ interpreted as the momentum and $\Th$ as the position. In this guise they are well-known: $X$ is the Hamiltonian of the harmonic oscillator, and $Y$ describes a particle in a linear potential in front of a wall. Here the quadratic form point of view is dictated by our aim to bound the expectation value $\braket\psih{Y\psih}$ for functions in the positive-time subspace, and corresponds in operator theoretic terms to taking the Dirichlet boundary condition at $t=0$. Note that this also corresponds to our discussion preceding \eqref{firstorderpsi}, which says that for the initial vectors $\psi$ satisfying our assumptions, we have $j\psi=\psih(0)=0$.  Both operators have purely discrete spectrum, because the potentials diverge at infinity. Moreover, the eigenvalues do not depend on the parameters $\tau,\varepsilon,\eta$, because of the unitary equivalence. To compute the bottom eigenvalue, we can thus set $\varepsilon=0,\tau=0,\eta=1$.

For $X$ we have the eigenvalues $x_n=(2n+1)$ (note that we left out a
conventional factor of $1/2$ from the definition of the oscillator Hamiltonian). The ground state is, of course, a Gaussian.

The eigenvalue equation for $Y$ becomes the differential equation
\begin{equation}\label{AiryEq}
    -\psi''(t)+t\psi(t)=y\psi(t),
\end{equation}
which is solved in terms of the Airy function $\Airy$ by
\begin{equation}\label{Airysol}
    \psi(t)= \Airy(t-y).
\end{equation}
Indeed, $\Airy$ is defined to be that solution of $-\Airy''(t)+t\Airy(t)=0$,
which is square integrable at $t\to+\infty$. The eigenvalue equation comes out
of the boundary condition: we must have $\psi(0)=\Airy(0-y)=0$, i.e., $(-y)$
must be a zero of $\Airy$ on the negative half axis. These zeros are
well-known and tabulated \cite[Table~9.9.1]{NISTfunctions}. The first two are
\begin{eqnarray}\label{AiryZero}
    Z_1&=&-y_0=- 2.33810\ 74104,
    \\
    Z_2&=&-y_1=- 4.08794\ 94441.
\end{eqnarray}

To get inequality \eqref{ETU} we choose $\varepsilon=\braket\psih{\Hh\psih}$
and $\tau=\expect\Th$. Taking expectations of the inequality $X\geq x_0\idty$
we find
\begin{equation}\label{delsum2}
    \eta^2\Delta\Hh^2+\eta^{-2}\Delta\Th^2\geq x_0=1.
\end{equation}
Minimizing this expression over $\eta$, i.e., setting
$\eta^2=\Delta\Th/\Delta\Hh$ we get $2\Delta\Th\,\Delta\Hh\geq1$ for the
canonical pair $\Th,\Hh$, and \eqref{ETU} as stated by
using \eqref{DelEHh}. We went through this well-known argument just to stress
the analogy with the proof of \eqref{EeTU}, for which the corresponding
expression using $Y$ reads
\begin{equation}\label{delsum1}
    \eta^2\Delta\Hh^2+\eta^{-1}\expect\Th\geq y_0.
\end{equation}
This time the minimum is attained at $\eta^3=\expect\Th/(2\Delta E^2)$, giving
\begin{equation}\label{delsum1a}
    3\left(\frac{\expect T \Delta\Hh^2}2\right)^{\frac23}\geq y_0.
\end{equation}
Together with \eqref{DelEHh} this gives \eqref{EeTU}.
It now remains to show \eqref{nearmin} for the almost minimal case. For this we have to first provide an elementary estimate, which quantitatively captures the heuristic idea that in a gapped ground state problem only states close to the ground state can have expectations close to the bottom eigenvalue.

\subsection{Lemma on gapped ground states}
Let $A$ be an operator, which has a lowest, non-degenerate eigenvalue $a_0$ with normalized eigenfunction $\phi_0$, so that the rest of the spectrum which lies in the half-axis $[a_1,\infty)$ with gap $a_1-a_0>0$. That is, we can write
\begin{eqnarray}\label{Agap}
    A&=&a_0\kettbra{\phi_0}+A_1
         \quad\mbox{with }\\
    A_1&\geq& a_1(\idty-\kettbra{\phi_0}).
\end{eqnarray}
Now suppose that for some state $\rho$ we have
\begin{equation}\label{nearbottom}
    \tr\rho A\leq a_0+\alpha,
\end{equation}
where $\alpha>0$. Then we claim that
\begin{equation}\label{gapestimate}
    \norm{\rho-\kettbra{\phi_0}}_1\leq2\sqrt{\frac\alpha{a_1-a_0}},
\end{equation}
where $\norm X_1=\tr\sqrt{X^*X}$ denotes the trace norm of the operator $X$.
The reason to write this in terms of density operators and trace norms (although we only need the case of a pure state $\rho=\kettbra\psih$ later on) is that in this form it is clearer that the bound is equivalent to
\begin{equation}\label{gapestimate1}
    \left|\tr(\rho S)-\braket{\phi_0}{S\phi_0}\right|\leq2\sqrt{\frac\alpha{a_1-a_0}},
\end{equation}
for all operators $S$ with $\norm S\leq1$.
Note that for any two density operators we have $\norm{\rho_1-\rho_2}\leq2$.
Hence the bound trivializes for $\alpha\geq a_1-a_0$, but gives some information as long as the expectation $a_0+\alpha$ lies in the gap.

For the proof we evaluate the inequality $$A\geq a_1\idty-(a_1-a_0)\kettbra{\phi_0}$$ with $\rho$, getting
$$ a_0+\alpha\geq\tr\rho A\geq a_1-(a_1-a_0)\braket{\phi_0}{\rho\phi_0},$$
which amounts to an estimate for the fidelity $\braket{\phi_0}{\rho\phi_0}$, namely
\begin{equation}\label{1-gapfides}
    1-\braket{\phi_0}{\rho\phi_0}\leq \frac\alpha{a_1-a_0} .
\end{equation}
The trace norm estimate follows from this by \cite[Lemma 4.1]{variazioni}.

\subsection{Almost minimal case}
Let us begin with the variance inequality. Of course, it is a standard exercise in almost every quantum physics textbook to show that the uncertainty relation holds with equality exactly for Gaussians. Much less well-known is the stability statement for this minimal case, which we prove here. It makes an almost equally simple exercise, so one might have expected to see it posed in quantum mechanics courses the world over. However, we could not find a reference for it.

So suppose that $\Delta T\Delta E\leq (1+\varepsilon)\sqrt p/2$. By \eqref{DelEHh} this implies $2\Delta\Th\Delta\Hh\leq(1+\varepsilon)$. The left hand side of this inequality is just the expectation of $X$ with the minimizing choices of constants. Hence the assumption \eqref{nearbottom} is satisfied and the estimate \eqref{gapestimate1} applies, which gives
\begin{equation}\label{stability2}
    \left|\tr(\rho S)-\braket{\phi_0}{S\phi_0}\right|\leq 2\sqrt{\frac\varepsilon{x_1-x_0}}=\sqrt{2\varepsilon}
\end{equation}
for any $S$ with $\norm S\leq1$. We choose $S$ as the multiplication operator with  $s(t)=\sign(\PP(t)-\Pmin(t))$.
Then
\begin{eqnarray}
  \norm{\PP-\Pmin}_1&=& \int_0^\infty\!\!dt\ \abs{\PP(t)-\Pmin(t)} \nonumber\\
     &=& \int_0^\infty\!\!dt\ s(t)\bigl(\PP(t)-\Pmin(t)\bigr) \nonumber\\
     &\leq&\sqrt{2\varepsilon}.
\end{eqnarray}

The argument for near equality of \eqref{EeTU} is completely analogous. Suppose that $\expect T\Delta E\leq(1+\varepsilon)c\sqrt p$, so
$\expect\Th\Delta\Hh\leq (1+\varepsilon)c$. Then the expectation of $Y$, which is the left hand side of \eqref{delsum1a} is by a factor $(1+\varepsilon)^{(2/3)}$ larger than the absolute minimum. We can hence apply the estimate with $\alpha=((1+\varepsilon)^{(2/3)}-1)y_0\leq2y_0\varepsilon/3$, which gives
\begin{eqnarray}\label{stability1}
    \norm{\PP-\Pmin}_1&\leq& \gamma\sqrt{\varepsilon}\\
    \mbox{with}\quad
    \gamma&=&2\sqrt{\frac{2y_0}{3(y_1-y_0)}}
             \approx1.888\ .
\end{eqnarray}

\vfill\strut

\end{document}